# Silicon Spin Qubit Control and Readout Circuits in 22nm FDSOI CMOS

Raffaele R. Severino, Michele Spasaro, and Domenico Zito

Dept. of Engineering, Electrical and Computer Engineering, Aarhus University, Finlandsgade 22, 8200 Aarhus N

*Abstract* — **This paper investigates the implementation of microwave and mm-wave integrated circuits for control and readout of electron/hole spin qubits, as elementary building blocks for future emerging quantum computing technologies. In particular, it summarizes the most relevant readout and control techniques of electron/hole spin qubits, addresses the feasibility and reports some preliminary simulation results of two blocks: transimpedance amplifier (TIA) and pulse generator (PG). The TIA exhibits a transimpedance gain of 108.5 dBΩ over a -3dB bandwidth of 18 GHz, with input-referred noise current spectral density of 0.89 pA/√Hz at 10 GHz. The PG provides a mm-wave sinusoidal pulse with a minimum duration time of 20 ps.**

*Keywords* — *Pulse generator, quantum dots, TIA, VCO*

## I. INTRODUCTION

Quantum computing (QC) supremacy [1] promises disruption in many fields, from drug discovery [2] to information security [3]. Several technologies and techniques have been explored so far, with a growing focus on developing scalable solid-state platforms. Once again silicon (Si) is emerging as a promising technology and pioneering works have already demonstrated high-fidelity operation with basic Si qubit devices [4, 5]. These breakthroughs stimulate the integration of Si electron/hole spin quantum bits (qubits) and circuits for spin control and readout together on the same chip, in commercial silicon-on-insulator (SOI) complementary metal oxide semiconductor (CMOS) foundry technology [6]. Qubits demonstrated so far require a temperature lower than 0.1 K, irrespective of their physical implementation [7, 8]. This prevents the co-integration of qubits and electronic circuits due to the difficult dc power dissipation at such extreme cryogenic temperatures. The multi-chip approach (e.g., [9]) may appear as the solution to this open challenge, but it is impractical for highly scalable QC platforms [8]. Fully depleted SOI (FDSOI) CMOS technology could allow qubit operation at a few Kelvins [10] and co-integration of qubits with control and readout circuits. This would be a major breakthrough, as it allows moving quantum technologies out from research laboratories to large-scale production. The EU H2020 project IQubits aims at demonstarting the feasibility of qubit operation at acceptable cryogenic temperatures, e.g., 3 K and above, and their co-integration with control and readout circuits in 22nm FDSOI CMOS.

This paper focuses on the preliminary design of the transimpedance amplifier (TIA) and pulse generator (PG), addressing the basic principles underlying the relatively high-T electron/hole spin qubit control and readout circuits, as a first step towards the implementation of the whole quantum processor. The paper is organized as follows. Section II summarizes some of the latest examples of silicon spin qubits. Section III reports the basic principles of qubit control and readout. Section IV addresses the design of the PG and TIA for qubit control and readout in 22nm FDSOI CMOS technology commercially available.

This work was supported in part by the European Commission through the European H2020 FET OPEN project IQubits (www.iqubits.eu, Grant Agreement N. 829005), and in part by the Poul Due Jensen Foundation.

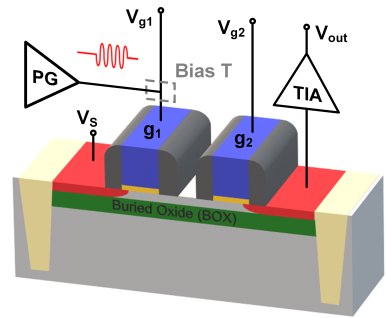

Fig. 1. FDSOI double quantum dot with control and readout circuits [10].

## II. SI SPIN QUBITS

Here, we summarize the main advantages and the most relevant approaches to implement Si spin qubits. Fidelity is a measure of how well a physical system implements an ideal quantum gate [9, 11] and depends on several factors, including the coherence time, i.e., the time scale over which the quantum superposition state holds phase coherence between basis states [12]. Long coherence times, together with scalability (small footprint), is likely the main advantage of spin qubits compared to others, e.g., super-conducting qubits [13]. For solid-state electron spin qubits, the coupling between the electron and nuclear spins is a major cause of decoherence. Compared to other semiconductors (e.g., III-V group), silicon has a sole stable isotope with non-zero nuclear spin (i.e., $^{29}$Si), which is rare in nature, and purified silicon (i.e., isotope $^{28}$Si) allows implementating qubits with longer coherence time due to the reduced interaction with nuclear spins in the host crystal. In $^{28}$Si devices, such single-qubit gates attained high fidelity of 99.9% and above [5]. Another significant advantage of silicon qubits is their compatibility with state-of-the-art integrated circuit (IC) manufacturing technologies. This is a key asset for implementing large array of identical objects, co-integrating qubits with classical electronics for control and readout. Single hole and electron spin qubits were recently implemented on SOI CMOS technology platforms [6, 10]. A way to realize Si qubits is to create a 2D electron gas in the proximity of a Si/SiGe junction. Depletion gates provide lateral confinement, tailoring the potential into a quantum dot (QD) [14]. A more compact solution is to exploit an accumulation field-effect gate to produce the confinement potential under a $Si/SiO_2$ interface [6, 10, 14]. Lateral confinement is obtained by patterning the Si active area. The device is similar to a MOSFET, but typically requires wider gate-source/gate-drain spacers. The gate voltage controls the number of charges confined in the region underneath the gate (i.e., Coulomb blockade). In principle, a single carrier can be isolated in the QD. The degeneracy between the spin states can be lifted by applying a static magnetic field $B_0$ (i.e., Zeeman splitting), enabling to encode information in the two spin states. The energy separation between the two levels is $E_Z = g\mu_B B_0$, being $g$ the $g$-factor and $\mu_B$ the Bohr magneton.

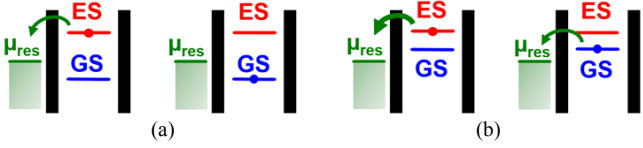

Fig. 2. Energy diagrams depicting two different methods for spin-to-charge conversion: (a) energy-selective and (b) tunnel-rate-selective readouts.

In FDSOI CMOS, the back gate in the Si substrate below the buried oxide (BOX) can be used to control the potential barrier between two adjacent devices, creating a double quantum dot (DQD) [6, 10]. The DQD, shown in Fig. 1 [10, 16], is essentially a two-gate n-MOSFET. A QD is formed in the undoped channel under each gate. In the 22nm FDSOI, the conventional spacers are sufficient to form potential barriers confining carriers between source and drain [10]. Gate oxide, BOX, and shallow trench isolation provide confinement along the other directions. In a DQD the information is encoded in singlet-triplet states [15], discussed in Section III. This relaxes the requirement of $E_Z$ being much larger than the thermal energy, enabling operation at a higher temperature and lower Larmor frequency ($\omega_L = E_Z/\hbar$).

## III. CONTROL AND REDOUT OF QUBITS

This section briefly reports some key aspects of qubit control and readout. Under the effect of a static field $B_0$ in the z-axis direction, an electron in a single QD acts as a two-level system. The electron spin precesses around the z-axis with the Larmor frequency $\omega_L = \gamma B_0$, being $\gamma$ the gyromagnetic ratio. If the system is perturbed with a magnetic field rotating at $\omega_L$ on the xy-plane, the spin-up probability of the electron in the QD varies periodically with the duration of the perturbation. The angular frequency $\omega_R$ of these so-called Rabi oscillations is named Rabi frequency and is proportional to the amplitude of the perturbation. In practice, a rotating field is approximated by an oscillating field with a frequency $\omega_L$ at resonance. A linear oscillation can be seen as the sum of two counterrotating motions, one resonating with the precessing magnetic moment and the other rotating in the opposite sense. In a reference frame rotating at $\omega_L$ with the precessing moment, the antiresonant component of the oscillating field rotates at $2\omega_L$. If $\omega_R \ll \omega_L$, these fast rotations can be neglected [17], assumption known as rotating wave approximation (RWA).

### A. ESR and EDSR

A first method to induce coherent spin rotations is the electron-spin resonance (ESR) [14]. The oscillating magnetic field $B(t)$ is generated by a large ac current $I(t)$ flowing in a microstrip line (ESR-line) deposited near the quantum dot. A second technique is the electric dipole-spin resonance (EDSR). The spin rotation is caused by an oscillating electric field, which is achieved applying an oscillating voltage $V(t)$ to a field-effect gate placed directly above the QD [14]. The electric field induces an orbital motion affecting the spin, if the spin-orbit coupling (SOC) is significant. Alternatively, when the SOC is weak, the same effect is obtained by embedding a micro-magnet close to the device, so that the oscillating particle perceives an oscillating magnetic field.

The relations $B(t) = \alpha_I I(t)$ and $B(t) = \alpha_V V(t)$ hold for ESR and EDSR, respectively. The conversion factors $\alpha_I$ and $\alpha_V$ depend on the geometrical and physical properties of the structures. They can be determined experimentally [17].

### B. Readout

Here, we discuss common techniques for Si spin qubit readout through spin-to-charge conversion: Energy-selective readout (ERO), tunnel-rate-selective readout (TR-RO), and readout based on spin blockade [15, 18]. The techniques correlate each possible spin state to a different charge state, which is determined through a charge detector, e.g., a single electron transistor (SET) or a quantum point contact (QPC).

ERO requires a QD coupled to a charge reservoir. Acting on the dot gate (plunger gate), the energy levels in the QD can be shifted relative to the electrochemical potential of the reservoir $\mu_{res}$. If the energy level of the excited spin state $|ES\rangle$ is higher than $\mu_{res}$ and that of the ground state $|GS\rangle$ is lower than $\mu_{res}$ as in Fig. 2(a), only an electron in $|ES\rangle$ can tunnel to the reservoir, while an electron in $|GS\rangle$ stays in the QD.

In TR-RO, an electron tunnels to a reservoir with a rate depending on the spin state. The concept is illustrated in Fig. 2(b). Assume that the tunnel rate $\Gamma_{ES}$ from $|ES\rangle$ to the reservoir exceeds the tunnel rate $\Gamma_{GS}$ from $|GS\rangle$. Initially, the energy levels of the two states are both set higher than $\mu_{res}$, so one electron can tunnel to the reservoir regardless of its spin state. After a time $\tau$ such that $\Gamma_{GS}^{-1} \gg \tau \gg \Gamma_{ES}^{-1}$, an electron in $|ES\rangle$ has probably tunneled out of the dot, while an electron in $|GS\rangle$ is still in the dot. The initial spin state is inferred counting the electrons in the dot. With readout schemes as ERO or TR-RO based on tunneling to the reservoir, the electron is lost in the reservoir after tunneling. In other schemes, the electron tunnels to a second dot, instead of the reservoir, so the electron is preserved and non-demolition measurements are possible.

In a DQD, as in Fig. 1, the information is encoded in a singlet-triplet scheme, instead of the Zeeman-split spin states of an electron in a single QD, and spin blockade is a viable readout technique [15]. Electrons can be transferred from one dot to the other by acting on the gate voltages to detune the electrochemical potentials of the dots. For two electrons in the same (say the right) QD, four spin states are possible: one singlet state $S(0,2)$ and three triplet states $T_0(0,2)$, $T_+(0,2)$, and $T_-(0,2)$. In the (1, 1) charge state, the two-electron states are also singlet and triplets, but with one electron per dot. Due to spin conservation in interdot transitions, the interdot coupling causes the (1, 1) singlet (triplet) states to hybridize with the corresponding (0, 2) singlet (triplet) states, and the avoided crossings of the singlets and triplets occur at different detunings. An external static magnetic field moves $T_+$ and $T_-$ far away, confining the relevant state space to S and $T_0$. The readout works as follows. The DQD is initialized in the (0, 1) configuration. Then, an electron is let into the left dot forming an $S(1,1)$ or a $T_0(1,1)$ state. $S(1,1)$ can be transformed into $S(0,2)$ under proper detuning. Instead, if a triplet is formed, the current is blocked until the electron spin relaxes. To discriminate between singlet and triplet states, the transition is detected through a charge sensor. Other readout techniques, such as gate reflectometry [7], are also possible.

### C. Example of Manipulation and Readout

An example of manipulation and readout of gate-defined Si DQD [19] is shown in Fig. 3. Fig. 3(a) illustrates the device, with two QDs formed below the gates L and R by depletion of a 2D electron gas accumulated at a Si/SiGe junction in the Si quantum well. Voltages on gates L and R are tuned to move the system across the charge stability diagram in Fig. 3(b). Gate M controls the singlet-triplet energy difference $J$.

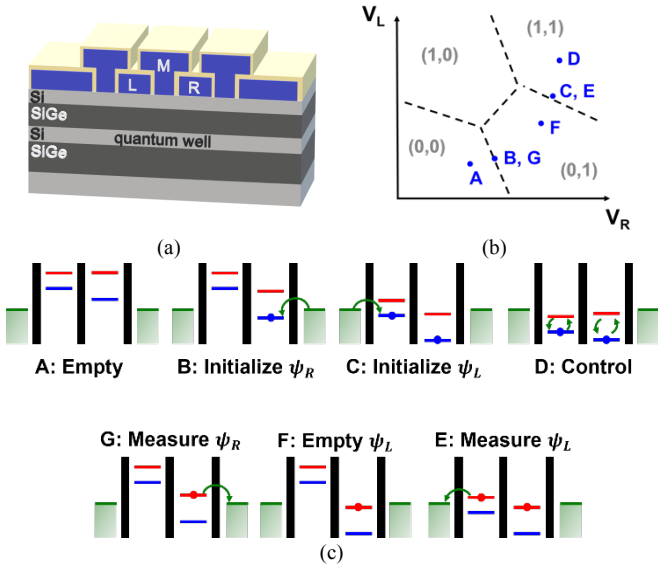

Fig. 3. Example of DQD manipulation and readout [19]. (a) Idealised structure. The energy levels of left and right dots are controlled through gates L and R, respectively. Gate M voltage tunes the coupling. (b) Charge stability diagram. Points A to G map the control/readout sequence. (c) Energy level configuration for each step of the control/readout sequence.

Fig. 3(c) illustrates the sequence of control/readout steps. Starting from A with charge state (0, 0), the device is initialized in the $|\downarrow\downarrow\rangle$ state by visiting B and C. With the system kept in the charge state (1, 1) in D, control pulses are applied to an ESR line, manipulating the spin. In order to measure the spin-up probability in the two QDs, the left QD undergoes ERO in E and emptied in F; next, the right QD is measured in G, then emptied, with the system returning in A.

## IV. MONOLITHIC QUBIT ICS

According to our vision, control and readout circuits for quantum computing should exhibit very low noise, provide wideband frequency operation, consume very low power and have a small form factor, so reaching the extreme performance requirements and assuring scalability to large qubit numbers.

As reported in Section I, the ambition of IQubits project is to demonstrate the feasibility of quantum bits operation at a few Kelvins, e.g. 3 K and possibly above. We consider the approach reported in [16]. An FDSOI DQD hosts the spin qubits; a PG is required for spin manipulation and the readout circuit is based on a TIA connected to the drain of the DQD, as shown in Fig. 1. Here, we carry out some considerations on the performance of control and readout circuits and investigate the implementation of PG and TIA. All circuit simulations are carried out considering the complete transistor models including the parasitics introduced by the backend of line, and inductor and capacitor models verified experimentally.

### A. Performance Considerations

The energy splitting between the two levels of the qubit should be at least as large as $k_B T$ [10], where $k_B$ is Boltzmann constant and $T$ is the temperature. Thereby, 3 K qubit operation requires minimum energy splitting of about 0.25 meV, corresponding to a Larmor frequency of about 60 GHz.

The evolution of a quantum system is captured by a unitary operator (U). The detailed physics of the real quantum system differs from the ideal quantum gate that it should implement. Moreover, noise and tolerances cause further discrepancies.

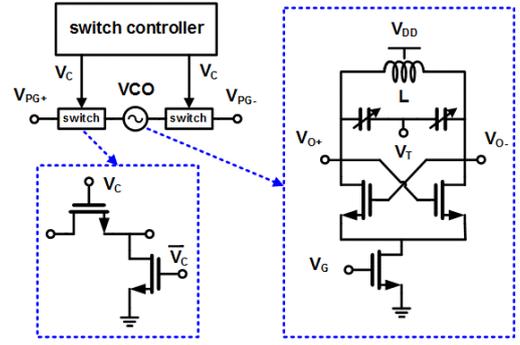

Fig. 4. Top-level schematic of the pulse generator.

The fidelity (F) of a quantum operation is a measure of the closeness between U and the time-evolution operator ($U_{id}$) of the ideal quantum gate. Also, infidelity (I) is defined as I = 1 – F. The performance of the electronic circuits affects F [17].

The phase noise (PN) of the voltage controlled oscillator (VCO) in the PG is a source of infidelity and [17] provides an approach to estimate roughly its infidelity contribution, assuming for simplicity a PN slope of -20 dB/dec. We consider a $\pi/2$-rotation qubit gate and a target infidelity contribution of $125 \times 10^{-6}$, mirroring the example of total error budget allocation to infidelity contributions provided in [17], which targeted a fidelity of 99.9% with $\omega_R \leq \omega_L/80$. For $\omega_R = \omega_L/80 = 750$ MHz, the acceptable inaccuracy ($\Delta$f) of the carrier frequency amounts to 11.8 MHz, corresponding to -74 dBc/Hz PN at 1 MHz from the carrier. For $\omega_R = \omega_L/5 = 12$ GHz, $\Delta$f amounts to 190 MHz, corresponding to -62 dBc/Hz at 1 MHz from the carrier. However, for $\omega_L/\omega_R = 5$, the RWA limits the fidelity to about 99.3%, as proved in [17].

So far, the highest values of $\omega_R$ reported for Si spin qubits are of the order of 100 MHz [7]. We consider $\omega_L/\omega_R$ ratios of 80 and 5, corresponding to maximum $\omega_R$ of 750 MHz and 12 GHz, and $\pi/2$ pulse durations of about 330 and 20 ps, respectively, with the latter being an extreme case. A 1.4% uncertainty ($\Delta$t) in pulse duration, amounting to 4.7 ps and 280 fs for $\omega_R$ of 750 MHz and 12 GHz, respectively, causes a further infidelity contribution of $125 \times 10^{-6}$.

Performance considerations for the TIA have been carried out in [10, 16]. The readout circuit must detect spin-selective electron/hole tunneling events resulting in peak currents in the order of 10 pA to 10 nA. In order to provide experimental proofs, TIA must amplify the current and convert it into an output voltage with a swing of few mV to be detected reliably by off-chip test equipment or processed by FPGAs. This corresponds to a transimpedance gain of 100-140 dBΩ, in case of off-chip test equipment at 50 Ω. Maximum output power transfer is also desirable.

The cryogenic characterisation of the 22nm FDSOI CMOS is reported in [10, 16]. It shows that the peak transconductance of n- and p-MOSFETs improves when moving from 300 K to 3.3 K, while the peak-$f_T$ and peak-$f_{MAX}$ current densities remain nearly constant. Passive devices exhibit lower losses at low temperature. All this allows us, in a first instance, to design circuits at 300 K expecting similar, or better, results at 3 K, somehow circumventing the limitations due to simulation convergence issues and unavailability of accurate models at cryogenic temperature. Also, the variation of the normalised transconductance below 77 K is quite limited [16], so we limit our analyses to 300 K and 77 K.

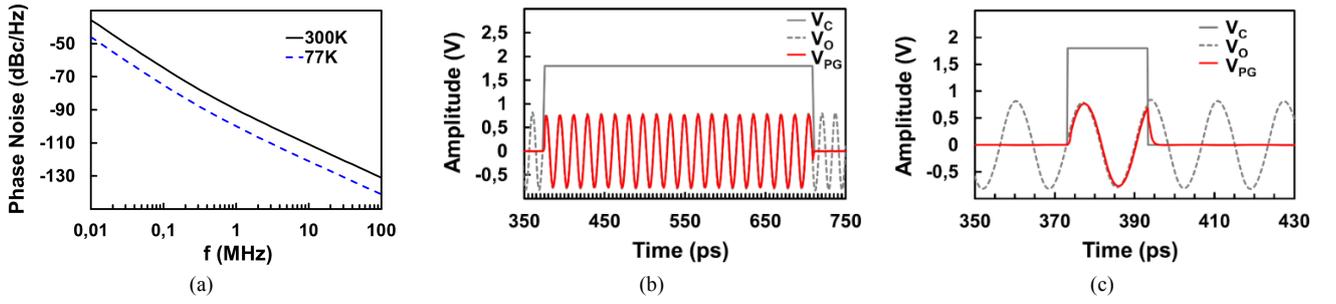

Fig. 5. (a) PN of the VCO at 300 K and 77 K. At 60 GHz, the PN amounts to -90 dBc/Hz at 1-MHz from the carrier, at 300 K and is reduced to -100 dBc/Hz at 77 K. Example of operation of the PG for (b) 330 ps π/2 pulse for 750 MHz Rabi frequency and (c) 20 ps π/2 pulse for 12 GHz Rabi frequency.

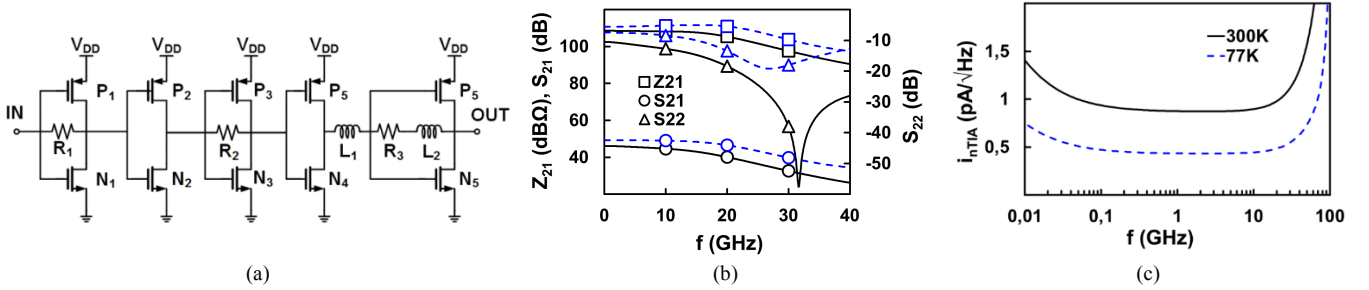

Fig. 6. (a) Schematic of the transimpedance amplifier. (b) Simulated $Z_{21}$, $S_{21}$, and $S_{22}$ parameters of the transimpedance amplifier at 300 K (black solid lines) and 77 K (blue dashed lines). (c) Input-referred noise current spectral density of the transimpedance amplifier at 300 K and 77 K.

## *B. Pulse Generator for Qubits Manipulation*

Our investigations show that the settling time required to generate a stationary wave with a VCO is about 200 ps, which is too long compared to the timing of qubit operation. For this reason we conceived a PG for qubit control and manipulation based on a VCO encapsulated within two switches, as shown in Fig. 4. The steady-state VCO output signal is fed into the quantum dot through the switch activated by a proper control signal, which provides the desired duration of the sinusoidal signal. As shown in Fig. 4, each switch consists of a pass MOSFET between the input and output nodes, and a second shunt MOSFET forcing the output to ground during the off-state, improving isolation. Owing to the switch, the VCO output signal is applied to the qubit only within the time interval with a minimum duration of about 20 ps.

In our study, we considered a cross-coupled differential pair VCO topology, as shown in Fig. 4. The LC tank consists of a symmetric center-tapped inductor of 114 pH and a quality factor of 22 at 60 GHz and a bank of n-MOSFET varactors. The tuning range spans about 8 GHz, from 57 to 65 GHz. $V_{DD}$ is 0.8 V. The PN is plotted in Fig. 5(a). At 300 K, PN = -90 dBc/Hz at 1 MHz from the 60 GHz carrier, well below -62 and -74 dBc/Hz emerging from fidelity considerations. At 77 K the PN is reduced to -100 dBc/Hz. The dc power consumption is lower than 3 mW at 300 K and rises to 3.8 mW at 77 K.

The PG operation and signals are reported in Figs. 5(b) and 5(c). The switches are activated by a pulse voltage $V_C$ of 1.8 V with rise- and fall-times equal to Δt/2 mirroring some of the fidelity considerations. The output voltage ($V_{PG}$) of the PG corresponding to a π/2 rotation is correctly generated at the Rabi frequencies of 750 MHz and 12 GHz, as shown in Figs. 5(b) and 5(c), respectively. The generated sinusoidal pulse is slightly longer due to the switch non-idealities.

## *C. Transimpedance Amplifier*

The TIA topology is shown in Fig 6(a); this is mutuated by the design in [10, 16]. However, here only two inductors of 400 pH with a quality factor of 20 at 13 GHz are adopted, leading to a more compact design for a small footprint and higher scalability within a multi-qubit readout processor. The TIA consists of an input stage implemented as a self-biased inverter, followed by two Cherry-Hooper stages. All transistors are biased at the peak-$f_{MAX}$ current density ($J_{fMAX}$ = 0.25mA/μm). MOSFETs in the input stage are sized for minimum noise contribution ($W_{N1}$ = 270 nm, $W_{P1}$ = 340 nm, $L_{N1} = L_{P1}$ = 20 nm). The next stages are designed to provide the best gain-bandwidth trade-off. The transimpedance gain ($Z_{21}$) and the S-parameters $S_{22}$ and $S_{21}$ at 300 K and 77 K are reported in Fig. 6(b). At 300 K, the TIA shows a $Z_{21}$ of 108.5 dBΩ and a -3dB bandwidth of 18 GHz. The output matching ($S_{22}$) is below -10 dB all over the bandwidth. The input-referred noise current spectral density ($i_{nTIA}$) is plotted in Fig. 6(c). $i_{nTIA}$ amounts to 0.89 pA/√Hz at 10 GHz. Simulations at 77 K show a $Z_{21}$ of 110.7 dBΩ with a -3dB bandwidth of 25 GHz while $i_{nTIA}$ reduces to 0.44 pA/√Hz. The dc power consumption amounts to 4.9 mW at 300 K and 5.5 mW at 77 K. Despite the output impedance matching and power consumption are slightly degraded, improvements of the gain and noise performances occur at low temperature, as expected.

## V. CONCLUSIONS

We have reported the most relevant techniques for control and readout of electron/hole spin qubits, and investigated the implementation of sinusoidal pulse generator and transimpedance amplifier integrated circuits in 22nm FDSOI CMOS technology. The circuits have shown the potential to operate with Si electron/hole-spin qubits with a Rabi frequency up to 12 GHz. In particular, the proposed pulse generator circuit can provide mm-wave sinusoidal pulses around 60 GHz for a π/2 rotation with minimum duration time of 20 ps. The transimpedence amplifier features a gain of 108.5 dBΩ over a -3dB bandwidth of 18 GHz, with an input-referred noise current spectral density of 0.89 pA/√Hz at 10 GHz. Future works are addressed to physical implementation and experimental verifications down to 3 Kelvins.